\documentclass[twocolumn,floatfix,prl]{revtex4}%
\usepackage{graphicx}%
\usepackage{amsmath}%
\usepackage{amsfonts}%
\usepackage{amssymb}
\usepackage{bm}
\usepackage{color}

\def\e{{\epsilon}}
\def\k{{ {\bm k} }}
\def\p{{ {\bm p} }}
\def\q{{ {\bm q} }}
\def\Q{{ {\bm Q} }}

\def\0{{ {\bm 0} }}
\def\w{{\omega}}
\def\a{{\alpha}}

\def\BI{{ B_{1g} }}
\def\BII{{ B_{2g} }}
\allowdisplaybreaks[4]

\begin{document}
\title{
Origin of diverse nematic orders in Fe-based superconductors:\\
45$^\circ$ rotated nematicity in AFe$_2$As$_2$ (A=Cs, Rb)
}
\author{
Seiichiro Onari and Hiroshi Kontani
}

\date{\today }

\begin{abstract}
The origin of diverse nematicity
and their order parameters in Fe-based
superconductors have been attracting increasing attention.
Recently, a new type of nematic order has been discovered
in heavily hole-doped ($n_d=5.5$) compound AFe$_2$As$_2$ (A=Cs, Rb).
The discovered nematicity has $\BII$ (=$d_{xy}$) symmetry, 
rotated by $45^\circ$ from the $\BI$ (=$d_{x^2-y^2}$) nematicity 
in usual compounds with $n_d\approx6$.
We predict that the ``nematic bond order'',
which is the symmetry-breaking of the correlated hopping,
is responsible for the $\BII$ nematic order in AFe$_2$As$_2$.
The Dirac pockets in AFe$_2$As$_2$
is essential to stabilize the $\BII$ bond order.
Both $\BI$ and $\BII$ nematicity
in A$_{1-x}$Ba$_x$Fe$_2$As$_2$ are naturally induced by 
the Aslamazov-Larkin many-body process, 
which describes the spin-fluctuation-driven charge instability.
The present study gives a great hint to control 
the nature of charge nematicity by modifying
the orbital character and the topology of the Fermi surface.

\end{abstract}

\address{
 Department of Physics, Nagoya University,
Furo-cho, Nagoya 464-8602, Japan. 
}
 

\sloppy

\maketitle

The electronic nematic state, which is the spontaneous
rotational symmetry breaking in the many-body electronic states,
appears in many Fe-based superconductors
\cite{ARPES}.
Above the structural transition temperature $T_{\rm S}$,
the electronic nematic susceptibility
develops divergently, observed as the 
softening of shear modulus $C_{66}$
\cite{Yoshizawa,Bohmer}, and
the enhancements of low-energy Raman spectrum
\cite{Gallais,Raman2}
and in-plane anisotropy of resistivity $\Delta\rho$
\cite{Fisher}.
The mechanism of nematicity and its order parameter
attract increasing attention,
as a key to understand the pairing mechanism of 
high-$T_{\rm c}$ superconductivity.
The intimate relationship between nematicity and magnetism 
has been discussed based on the spin-nematic scenarios
\cite{Fernandes,DHLee,QSi,Valenti,Fang,Fernandes-review,C4,Khasanov}
and the orbital/charge-order scenarios
\cite{Kruger,PP,WKu,Onari-SCVC,Onari-SCVCS,Onari-form,FeSe-Yamakawa,Text-SCVC,JP,Fanfarillo,Chubukov-RG}.


Beyond the initial expectations,
Fe-based superconductors exhibit
very rich phase diagrams with nematicity and magnetism.
In FeSe, for example, the nematic order does not 
accompany the magnetism at ambient pressure,
whereas this nonmagnetic nematic phase is suppressed and  
replaced with the SDW phase by applying pressure
\cite{FeSe-P1,FeSe-P2}.
This phase diagram is understood in terms of the 
orbital-order scenario by assuming the pressure-induced 
$d_{xy}$-orbital hole-pocket
\cite{FeSe-P3}.
In the orbital/charge-order scenario, the orbital/charge order is driven
by the spin fluctuations, due to the Aslamazov-Larkin (AL) vertex
correction (VC) that describes 
the charge-spin mode coupling.
The significance of the AL process
has been clarified by several theoretical studies, especially 
by renormalization group studies
\cite{Tsuchiizu-Ru1,Tsuchiizu-Cu,Tsuchiizu-CDW,Chubukov-RG,Chubukov-RG2,Schmalian}.
However, the origin of the diverse electronic states 
associated with charge, orbital and spin degree of freedoms
is not fully understood.

Until recently, all the discovered nematic orders 
in Fe-based superconductors
have $\BI$ (=$d_{x^2-y^2}$) symmetry, along the nearest Fe-Fe direction.
Recently, however, nematic order/fluctuation with 
$\BII$ (=$d_{xy}$) symmetry,
rotated by $45^\circ$ from the conventional $\BI$ nematicity,
has been discovered in heavily hole-doped ($n_d=5.5$) compound 
AFe$_2$As$_2$ (A=Cs, Rb).
Strong $\BII$ nematic fluctuations and static order
have been discovered by the NMR study 
\cite{CsFe2As2-nematic},
the quasiparticle-interference by STM
\cite{RbFe2As2-nematic},
and the measurement of in-plane anisotropy of resistivity
\cite{Shibauchi-B2g}
in RbFe$_2$As$_2$ ($T_c\sim 2.5$K) 
and CsFe$_2$As$_2$ ($T_c\sim 1.8$K).
No SDW transition is observed in both 
compounds down to $T_{\rm c}$.
\cite{CsFe2As2-HF,Shibauchi-B2g}.
Surprisingly, both $\BI$ and $\BII$ nematic transitions
are observed in Y-based 
\cite{Sato-BIg}
and Hg-based 
\cite{Murayama-BIIg}
cuprate superconductors,
respectively, at the pseudogap temperature $T^*$.
Theoretical studies of nematicity in cuprates
have been performed by many authors
\cite{Sachdev,Fradkin,Metzner,Metzner2,Yamakawa-c,PLee,Pepin,Kawaguchi-Cu,Tsuchiizu-Cu}.
The discovery of unexpected $\BII$ nematicity 
in both Fe-based and cuprate superconductors
puts a severe constraint on the mechanism of nematicity.

In this paper, 
to reveal the origin of the $\BII$ nematicity,
we study the spin-fluctuation-driven charge nematicity
in AFe$_2$As$_2$ by considering the higher-order VCs.
We predict that the ``nematic bond order'',
given by the symmetry-breaking in the 
$d_{xy}$-orbital correlated hopping,
is responsible for the $\BII$ nematic order in AFe$_2$As$_2$.
The Dirac pockets around X,Y points play essential role on 
the $B_{2g}$ bond order.
With electron-doping, it is predicted that the $\BII$ 
nematicity changes to the conventional
$\BI$ nematicity at the Lifshitz transition point, 
at which two Dirac pockets merge
into one electron Fermi surface (FS).
The diverse nematicity 
in A$_{1-x}$Ba$_x$Fe$_2$As$_2$ is naturally understood
since the charge nematicity caused by the AL-VCs
is sensitive to orbital character and topology of the FS.
The present study gives a great hint to control the nature of
nematicity in Fe-based superconductors.



\begin{figure}[!htb]
\includegraphics[width=.99\linewidth]{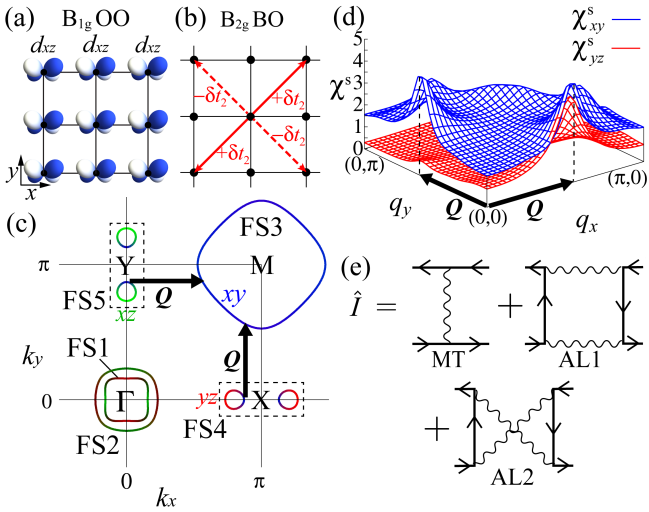}
\caption{
Schematic pictures of (a) $\BI$ orbital order (OO),
and (b) $\BII$ bond order (BO).
(c) FSs of the CsFe$_2$As$_2$ model in unfolded zone.
The colors green, red and blue correspond to 
orbitals 2, 3 and 4, respectively.
Each arrow denotes the significant 
intra-$d_{xy}$-orbital nesting vector $\bm{Q}=(0.53\pi,0)$.
(d) $\q$ dependences of 
 $\chi^{s}_{xy}(\q,0)$ and $\chi^{s}_{yz}(\q,0)$ given by the RPA.
(e) Feynman diagrams of the irreducible
four-point vertex $\hat{I}$. The wavy line is the
fluctuation-mediated interaction $\hat{V}^{s,c}$
}
\label{fig:FS}
\end{figure}

First, we introduce the nematic order parameters.
Figure \ref{fig:FS} (a) shows B$_{1g}$ nematic states 
due to orbital order ($n_{xz}\ne n_{yz}$).
Here, the ($x,y$) axes are along the nearest Fe-Fe directions.
The orbital order is the origin of the B$_{1g}$ nematicity in Fe-based superconductors.
Figure \ref{fig:FS} (b) shows B$_{2g}$ nematic state given by the next-nearest-neighbor (NNN)
bond order,
which corresponds to the modulation of
the NNN correlated hopping $\delta t_2$. 
We propose that the B$_{2g}$ bond order is the origin of
the B$_{2g}$ nematicity in AFe$_2$As$_2$,
which has not been discussed in previous theoretical studies
\cite{JP,Metzner,Metzner2,Sachdev}.

We analyze the following two-dimensional 
eight-orbital $d$-$p$ Hubbard model with parameter $r$
\cite{Onari-form}:
\begin{eqnarray}
H_{\rm M}(r)=H^0+rH^U, \ \ \
\label{eqn:Ham}
\end{eqnarray}
where 
$H^0$ is the unfolded tight-binding model derived from
the first-principles calculation for CsFe$_2$As$_2$,
which we introduce in the Supplemental Material (SM) A \cite{SM}.
$H^U$ is the first-principles screened Coulomb potential
for $d$-electrons in BaFe$_2$As$_2$ \cite{Arita}.
Figure \ref{fig:FS}(c) shows the Fermi surfaces (FSs):
The hole FS around M point (FS3) composed of $d_{xy}$-orbital is large,
while the Dirac pockets near X and Y points (FS4,5) are small.
The arrows denote the most important 
intra-$d_{xy}$-orbital nesting vectors.
Below, we denote the five $d$-orbital
$d_{3z^2-r^2}$, $d_{xz}$, $d_{yz}$, $d_{xy}$, $d_{x^2-y^2}$ as $l=1,2,3,4,5$.


We calculate the spin (charge) susceptibilities ${\hat \chi}^{s(c)}(q)$
for $q=(\q,\w_m=2m\pi T)$ based on the random-phase-approximation (RPA).
The spin Stoner factor $\alpha_{s}$ is given by
the maximum eigenvalue of $\hat{\Gamma}^{s}\hat{\chi}^0(\bm{q},0)$, where
${\hat \Gamma}^{s(c)}$ is the bare Coulomb interaction 
for the spin (charge) channel,
and $\hat{\chi}^0$ is the irreducible susceptibilities
given by the Green function without self-energy 
${\hat G}(k)=[(i\e_n-\mu){\hat1}-{\hat{h}}^0(\k)]^{-1}$ 
for $k=[\k,\e_n=(2n+1)\pi T]$.
Here, ${\hat{h}}^0(\k)$ is the matrix expression of $H^0$ 
and $\mu$ is the chemical potential.
Details of $\hat{\Gamma}^{s(c)}$, $\hat{\chi}^{s(c)}(q)$, 
and $\hat{\chi}^0(q)$ are explained in the SM A \cite{SM}.
We use $N=64\times64$ $\k$-meshes and $512$ Matsubara frequencies,
and fix the parameters $r=0.30$ and $T=0.03$ eV unless otherwise noted. 
Figure \ref{fig:FS}(d) shows the obtained spin susceptibility
$\chi^{s}_{xy(yz)}(\q,0)\equiv\chi^{s}_{l,l;l,l}(\q,0)$ 
with $l=4$ ($l=3$) at $\alpha_s=0.93$. 
$\chi^{s}_{xy}$ is enlarged due to the intra-$d_{xy}$-orbital nesting,
and it has the largest peak at $\q=\bm{Q}=(0.53\pi,0)$.
In contrast, $\chi^{s}_{yz}$ is small since 
the intra-$d_{yz}$-orbital nesting is bad.
Note that $\chi^{s}_{xy}\le\chi^{s}_{yz}$ in LaFeAsO, BaFe$_2$As$_2$,
and FeSe since two Dirac pockets (FS4 and FS5) 
merge into a usual electron pocket for $n_d\sim6.0$.

Hereafter, we study the symmetry-breaking in the self-energy 
($\Delta\hat{\Sigma}$) based on the 
density-wave (DW) equation introduced in Ref. \cite{Onari-form}.
We calculate both momentum- and orbital-dependences of 
$\Delta \Sigma_{l,l'}^\q(k)$ self-consistently 
in order to analyze both orbital order and bond order on equal footing.
To find the wavevector $\q$ of the DW state,
we solve the following linearized DW equation:
%

\begin{eqnarray}
\lambda_\q \Delta{\hat\Sigma}^\q(k)= \frac{T}{N}
\sum_{k'} {\hat K}^{\bm{q}}(k,k')\Delta{\hat \Sigma}^\q(k'),
\label{eqn:linearized}    
\end{eqnarray}
where $\lambda_\q$ is the eigenvalue for the DW equation.
The DW with wavevector $\q$ appears when $\lambda_\q=1$, and
the eigenvector $\Delta\hat{\Sigma}^\q(k)$ gives the DW form factor.
The kernel function
$\hat{K}^{\bm{q}}(k,k')$ \cite{Kawaguchi-Cu} is given by
\begin{equation}
{\hat K}^{\bm{q}}(k,k')={\hat I}^{\bm{q}}(k,k'){\hat g}^{\bm{q}}(k'),
\label{eqn:K}
\end{equation}
where 
$g^{\bm{q}}_{l,l';m,m'}(k)\equiv
G_{l,m}\left(k+\frac{\bm{q}}{2}\right)G_{m',l'}\left(k-\frac{\bm{q}}{2}\right)$,
and $\hat{I}^{\bm{q}}(k,k')$ is the irreducible four-point vertex.
It is given by the
Ward identity $\hat{I}=\delta {\hat \Sigma}/\delta {\hat G}$,
where $\hat\Sigma$ is one-loop self-energy
\cite{conserving}.
The Feynman diagram of $\hat{I}^{\bm{q}}$ is shown 
in Fig. \ref{fig:FS} (e):
The first diagram corresponds to the Maki-Thompson (MT) term,
and the second and the third diagrams are AL1 and AL2 terms, respectively.
Its analytic expression is given in the SM A\cite{SM}.
Near the magnetic criticality, the charge-channel interaction 
due to the AL terms is strongly enhanced
in proportion to $\sum_p\{\chi^s(\p,0)\}^2$,
which is proportional to $\chi^s(\bm{Q},0)$
in two-dimensional systems.
For this reason, the AL terms cause the 
spin-fluctuation-driven charge nematic order
\cite{Onari-SCVC,Onari-form,Tsuchiizu-Cu,Tsuchiizu-Ru1,Tsuchiizu-CDW}.

The Hartree-Fock (HF) term,
which is the first order term with respect to ${\hat{\Gamma}}^{s,c}$, 
is included in the MT term.
As well-known, the HF term suppresses conventional 
charge DW order ($\Delta\Sigma=$const), whereas 
both $\BI$ and $\BII$ bond orders are not suppressed.
Here, we drop the $\e_n$-dependence of 
$\Delta{\hat \Sigma}^\q(k)$ by the analytic continuation 
($\e_n \rightarrow \e$) and putting $\e=0$
 \cite{Onari-form}. 
This approximation leads to slight overestimation of $\lambda_\q$.

\begin{figure}[t]
\includegraphics[width=.99\linewidth]{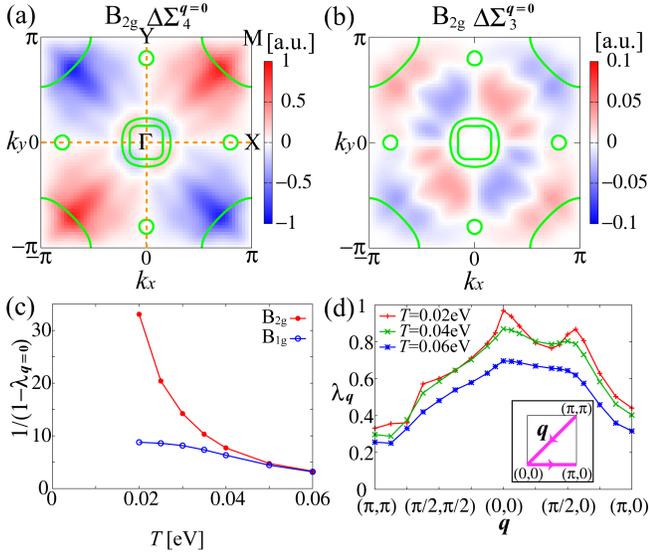}
\caption{
(a,b) $\BII$ symmetry form factors at $\q={\bm 0}$
obtained as the largest eigenvalue.
The primary form factor on $d_{xy}$ orbital,
$\Delta\Sigma_{4}^{\bm{0}}\propto \sin k_x \sin k_y$,
gives the bond order.
Orange dotted lines represent the symmetry nodes.
(c) The strengths of nematic fluctuations $1/(1-\lambda_{\bm{q}=\bm{0}})$ for
 $\BII$ and $\BI$ symmetries as a function of $T$.
(d) $\bm{q}$ dependences of the maximum eigenvalue 
at $T=0.02$eV, $0.04$eV, and $0.06$eV. 
}
\label{fig:Sigma}
\end{figure}


Figures \ref{fig:Sigma}(a) and \ref{fig:Sigma}(b) show 
the obtained form factors at $\q={\bm 0}$,
$\Delta\Sigma^{\bm{0}}_{4}(\k)\equiv\Delta\Sigma^{\bm{0}}_{4,4}(\k)$ and
$\Delta\Sigma^{\bm{0}}_{3}(\k)\equiv\Delta\Sigma^{\bm{0}}_{3,3}(\k)$,
for the largest eigenvalue $\lambda=0.93$.
(The absolute value of $\Delta\hat{\Sigma}^{\q}$ is meaningless.)
The obtained form factor has $\BII$-symmetry
since the symmetry relation
$\Delta\Sigma^{\bm{0}}_{4}(k_x,k_y)\propto \sin k_x \sin k_y$ holds.
The relation $|\Delta\Sigma_{xy}| \gg |\Delta\Sigma_{yz(xz)}|$
means that the primary nematic order is the 
``next-nearest-neighbor bond order for $d_{xy}$ orbital'',
which is shown in Fig. \ref{fig:FS} (b).
The obtained $\BII$ bond order is consistent with the
experimental $\BII$ nematicity
in AFe$_2$As$_2$
\cite{CsFe2As2-nematic,Shibauchi-B2g,RbFe2As2-nematic}.
The second largest eigenvalue $\lambda=0.88$ corresponds 
to the $\BI$ nematic bond order,
details of which we explain in the SM B \cite{SM}.

As explained in the SM C \cite{SM}, 
the nematic susceptibility with respect to
the form factor $\Delta\hat{\Sigma}^{\bm{q}}$ is given as
${\hat\chi}^{\Delta\Sigma}(\q) \propto (1-\lambda_{\bm{q}})^{-1}$
that diverges at $\lambda_{\bm{q}}=1$.
Figure \ref{fig:Sigma}(c) shows the $T$ dependences of
$(1-\lambda_{\bm{0}})^{-1}$
for both $\BII$ and $\BI$ symmetry solutions.
We see that $(1-\lambda_{\bm{0}})^{-1}$
for the $\BII$ symmetry shows the Curie-Weiss behavior
and dominates over that for the $\BI$ symmetry. 
These results are consistent with the
experimental nematic susceptibility \cite{CsFe2As2-nematic,Shibauchi-B2g}.
In Fig. \ref{fig:Sigma}(d), we show the $\q$ dependences of the largest
eigenvalue at $T=0.02$eV, $0.04$eV, and $0.06$eV. 
It is confirmed that the nematic susceptibility 
actually has the maximum peak at $\q=\bm{0}$, and the symmetry of form
factor is $\BII$.

\begin{figure}[!htb]
\includegraphics[width=.99\linewidth]{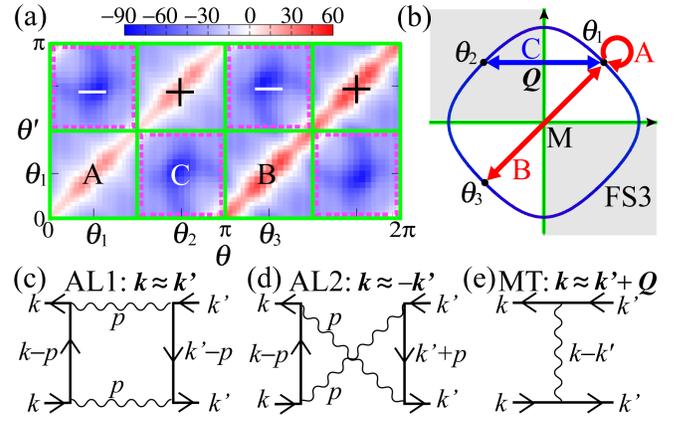}
\caption{
(a) $K_{\rm FS3}(\theta,\theta')$ on FS3
given by all vertex terms.
The green lines denote the $\BII$ symmetry nodes.
A, B, and C represent the pairs of Fermi points $(\theta_1,\theta_1)$,
 ($\theta_3,\theta_1$), and $(\theta_2,\theta_1)$, respectively: $\theta_1\equiv\pi/4$, $\theta_2\equiv3\pi/4$, and $\theta_3\equiv5\pi/4$.
(b) $B_{2g}$ symmetry order ($\Delta\Sigma(\k)\propto \sin k_x \sin k_y$)
driven by attractive (repulsive) interaction for pairs A and B (pair C).
(c,d,e) $\hat{I}^{\bm{0}}(k,k')$ given by AL1 term, AL2 term, and MT term.
Two AL terms give strong attractive interaction for $(\k,\pm\k)$, 
shown as red line regions in panel (a).
The MT gives repulsive interaction for pair C,
due to spin fluctuations at ${\bm Q}\approx(0.5\pi,0)$.
}
\label{fig:Kernel}
\end{figure}

In order to understand the origin of the $\BII$ nematic bond order,
we analyze the momentum-dependence of the kernel function
for $d_{xy}$ orbital.
Figure \ref{fig:Kernel} (a) shows 
$K_{\rm FS3}(\theta,\theta')\equiv
T\sum_{n'} K^{\bm{0}}_{4,4;4,4}
(\k(\theta),\e_n,\k(\theta'),\e_{n'})|_{{\e_n}\rightarrow0}$
given by the summation of the AL1, AL2, and MT terms on the FS3.
Here, $\theta$ and $\theta'$ denote the azimuthal angles (from the M point) of 
$\k$ and $\k'$ on the FS3, respectively.
Now, we define the pairs of Fermi points A$=(\theta_1,\theta_1)$,
 B$=(\theta_3,\theta_1)$, and C$=(\theta_2,\theta_1)$, where
$\theta_1\equiv\pi/4$, $\theta_2\equiv3\pi/4$, and
$\theta_3\equiv5\pi/4$.
For these pairs $K_{\rm FS3}(\theta,\theta')$ becomes
large in magnitude. The green lines denote the nodes of 
$\BII$ symmetry $(\theta,\theta'=\frac{\pi}{2}n)$. 
The positive $K_{\rm FS3}(\theta,\theta')$ for the pairs A and B give
attractive interactions between the same $(\k_1,\k_1)$ and the opposite
$(-\k_1,\k_1)$ momenta in Eq. (\ref{eqn:linearized}), respectively,
where $\k_i\equiv\k(\theta_i)$ $(i=1,2,3)$.
On the other hand, the negative $K_{\rm
FS3}(\theta,\theta')$ for the pair C gives the repulsive
interaction between $(\k_2,\k_1)$.
As we show in
Fig. \ref{fig:Kernel} (b),
this checkerboard-type sign structure of
$K_{\rm FS3}(\theta,\theta')$, which is positive (negative) 
for pairs A and B (pair C),
favors the $\BII$ symmetry bond order
$\Delta\Sigma^{\bm{0}}_4(\k)\propto \sin k_x\sin k_y$.

We briefly explain the microscopic origin of the checkerboard-type sign structure in 
$K_{\rm FS3}(\theta,\theta')$.
The positive $K_{\rm FS3}(\theta,\theta')$ along $\theta'=\theta$ in
Fig. \ref{fig:Kernel} (a) (including the pair A) originates from
the AL1 term, since the particle-hole channel $\phi_{\mbox{p-h}}\equiv
T\sum_p G_{4,4}(k-p)G_{4,4}(k'-p)$ shown in Fig. \ref{fig:Kernel}
(c) takes large positive value for $\k'=\k$, as we explain in
the SM D \cite{SM}.
Also, the positive $K_{\rm FS3}(\theta,\theta')$ along $\theta'=\theta+\pi$ (including the pair B)
originates from the AL2 term, since the particle-particle (Cooper)
channel $\phi_{\rm p\mbox{-}p}\equiv T\sum_{p}G_{4,4}(k-p)G_{4,4}(k'+p)$
shown in Fig. \ref{fig:Kernel} (d) takes large positive value for $\k'=-\k$.
On the other hand, the negative $K_{\rm FS3}(\theta_2,\theta_1)$ at the pair C 
stems from the MT term in Fig. \ref{fig:Kernel} (e).
This is because $\hat{V}^s(k-k') \propto \hat{\chi}^s(k-k')$ in the MT
term becomes maximum for $(\k,\k')=(\k_2,\k_1)$ 
since $\k_2-\k_1$ coincides with the nesting vector $\Q$.

To summarize, 
both $\BI$ and $\BII$ nematicities can be induced by the AL terms, 
since they give attractive interaction for both 
$\theta \approx \theta'$ and $\theta \approx \theta'+\pi$.
In fact, both the nematic susceptibilities $(1-\lambda_{\q})^{-1}$ 
for the $\BI$ and the $\BII$ increase
as shown in Fig. \ref{fig:Sigma} (c), consistently with recent experiment\cite{Shibauchi-B2g}.
In the present model with spin fluctuations at ${\bm Q}\approx(0.5\pi,0)$, 
the $\BII$ nematic order is assisted by the MT term.
The magnitude of the AL kernel function
dominates over that of the MT kernel function
as we explain in SM D \cite{SM}.
For this reason, the eigenvalue of the DW equation $\lambda_\q$
can be larger than that of the Eliashberg gap equation,
in which the kernel contains only the MT term \cite{Norman}.
We predict that the 
$\BII$ nematicity is closely tied to the Dirac pockets,
which give the main spin fluctuations in AFe$_2$As$_2$.

\begin{figure}[!htb]
\includegraphics[width=.99\linewidth]{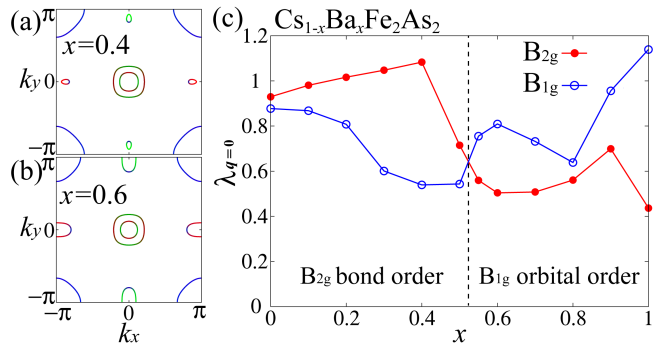}
\caption{
(a) FSs for $x=0.4$ and (b) FSs for $x=0.6$
in the Cs$_{1-x}$Ba$_x$Fe$_2$As$_2$ model.
(c) $x$ dependences of $\lambda$ for $\BII$- and $\BI$ symmetry 
obtained in the Cs$_{1-x}$Ba$_x$Fe$_2$As$_2$ model. 
}
\label{fig:xdep}
\end{figure}

Here, we discuss the doping-dependence of the nematicity:
We introduce reliable model Hamiltonian for
Cs$_{1-x}$Ba$_x$Fe$_2$As$_2$, by interpolating between 
CsFe$_2$As$_2$ model and BaFe$_2$As$_2$ model with the ratio $1-x:x$.
With increasing $x$, the FSs with four Dirac pockets
in Fig. \ref{fig:xdep}(a) for $x=0.4$ change to the 
FSs with two electron pockets in Fig. \ref{fig:xdep}(b) for $x=0.6$.
In this model, 
the Lifshitz transition occurs at $x_c\approx0.5$.

Figure \ref{fig:xdep}(c) shows $x$ dependences of 
$\lambda_{\bm{q}=\bm{0}}$ for
the $\BII$ and the $\BI$ symmetries in the
Cs$_{1-x}$Ba$_x$Fe$_2$As$_2$ model, in which
value of $r$ is fixed to 0.30.
For $x<x_c$,
the $\BII$ bond order $\pm\delta t_2$ shown in Fig. \ref{fig:FS} (b)
is dominant over the $\BI$ orbital order, 
since the former is driven by strong 
spin fluctuations in $d_{xy}$ orbital.
For $x>x_c$, the $\BI$ orbital order $n_{xz}\ne n_{yz}$
in Fig. \ref{fig:FS}(a) becomes dominant,
because of the strong spin fluctuations in $d_{xz,yz}$ orbitals
due to the nesting between electron- and hole-FSs
 \cite{Onari-SCVC,Onari-SCVCS,FeSe-Yamakawa},
as we briefly explain in the SM E \cite{SM}.
Thus, the present theory 
naturally explains
both the $\BI$ nematicity in non-doped $(n_d\approx 6)$
systems and $\BII$ nematicity in heavily
hole-doped compounds in a unified way, 
by focusing on the impact of the Lifshitz transition.

\begin{figure}[!htb]
\includegraphics[width=.9\linewidth]{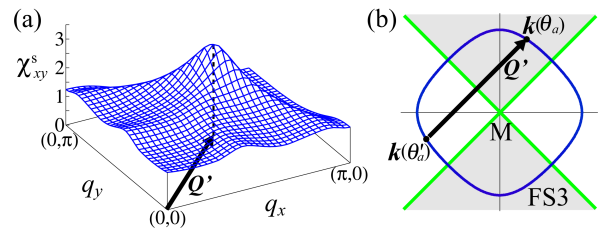}
\caption{
(a) $\chi^s_{xy}(q)$ for $r=0.36$ ($\a_S=0.90$)
given by dropping the contribution from the Dirac pockets.
(b) $B_{1g}$ nematic order with green symmetry nodes and gray negative
 region on FS3 due to the intra-FS nesting $\bm{Q}'\approx(0.6\pi,0.6\pi)$.
}
\label{fig:Dirac}
\end{figure}

The sudden decrease of $\lambda_{{\bm0}}^{\rm B_{2g}}$
at the Lifshitz transition point in Fig. \ref{fig:xdep} (c)
indicates that the Dirac pockets are essential for the 
$B_{2g}$ nematicity, in spite of their small size.
To verify this, we calculate $\chi^s_{xy}(q)$ 
by dropping the contribution from the rectangular areas
around X,Y points shown in Fig. \ref{fig:FS} (c):
Then, as shown in Fig. \ref{fig:Dirac} (a),
the peak at $\Q=(0.53\pi,0)$ of $\chi^s_{xy}(q)$ in Fig. \ref{fig:FS} (d)
shifts to $\Q'=(0.56\pi,0.56\pi)$, which is the 
intra-FS3 nesting vector.
In this case, $K_{\rm FS3}(\theta,\theta')$ due to MT term
takes large negative
value for $\theta\approx\theta_a$ and $\theta'\approx \theta_a'$
in Fig. \ref{fig:Dirac} (b),
and therefore $B_{1g}$ bond order emerges:
$\lambda_{\bm{0}}^{B_{1g}}=0.82$ and $\lambda_{\bm{0}}^{B_{2g}}=0.77$.
To summarize, the $\BII$ nematicity in AFe$_2$As$_2$ 
is closely tied to the emergence of the Dirac pockets
at the Lifshitz transition.
Thus, we can control the nematicity 
by changing the topology and orbital character of the FSs.

Recently, the $\BII$ vestigial nematic order 
has been proposed in Ref.  \cite{Fernandes-B2g,Si-B2g}
based on the real-space picture,
whereas the double stripe magnetism ($\q=(\pi/2,\pi/2)$)
has not been observed yet.
Thus, it is an important future issue to determine 
the mechanism of $\BII$ nematicity.


In summary, 
we studied the rich variety of nematic orders realized in
$A_{1-x}$Ba$_x$Fe$_2$As$_2$ (A=Cs, Rb)
by solving the DW equation with AL- and MT-VCs .
At $x=0$, the $\BII$ bond order 
is driven by the spin fluctuations in $d_{xy}$ orbital.
With increasing $x$, the $\BII$ nematicity 
suddenly changes to $\BI$ orbital nematicity ($n_{xz}-n_{yz}$)
at the Lifshitz transition point,
consistently with recent experiment
\cite{Shibauchi-B2g}.
Both the FS orbital character and the FS topology
are key ingredients not only to understand the diverse nematicity, 
but also to control the nature of nematicity in Fe-based superconductors.
The present theory will give useful hints to understand
recently-discovered rich nematic orders
in cuprate superconductors
\cite{Sato-BIg,Murayama-BIIg}.

We stress that the present DW equations
satisfy the criteria of the 
``conserving approximation (CA)'' 
by introducing the self-energy in $G$'s
\cite{Baym1,Baym2,ROP}. 
The great merit of the CA is that the 
macroscopic conservation laws are satisfied rigorously.
This merit is important to avoid unphysical results.
In the SM F \cite{SM}, we improve the present theory 
within the framework of the CA, 
by introducing the self-energy given by the
fluctuation-exchange (FLEX) approximation.
The obtained $\q$-dependences of $\lambda_\q$ and $\BII$ symmetry 
form factor are essentially similar to Fig. \ref{fig:Sigma}.
Thus, the main results of the present study are justified
within the framework of the CA.

\acknowledgements
We are grateful to Y. Matsuda, T. Shibauchi and Y. Yamakawa
for useful discussions.
This work was supported by Grant-in-Aid for Scientific Research from 
the Ministry of Education, Culture, Sports, Science, and Technology, Japan.




\clearpage

\makeatletter
\renewcommand{\thefigure}{S\arabic{figure}}
\renewcommand{\theequation}{S\arabic{equation}}
\makeatother
\setcounter{figure}{0}
\setcounter{equation}{0}
\setcounter{page}{1}
\setcounter{section}{1}

\begin{widetext}
\begin{center}
{\bf 
[Supplementary Material] \\
Origin of diverse nematic orders in Fe-based superconductors:\\
45$^\circ$ rotated nematicity in AFe$_2$As$_2$ (A=Cs, Rb)
}%
\end{center}

\begin{center}
Seiichiro Onari and Hiroshi Kontani
\end{center}

\begin{center}
\textit{Department of Physics, Nagoya University, Nagoya 464-8602, Japan}
\end{center}

\end{widetext}
\subsection{A: Eight-orbital models for AFe$_2$As$_2$ and BaFe$_2$As$_2$}

\begin{figure}[!htb]
\includegraphics[width=.99\linewidth]{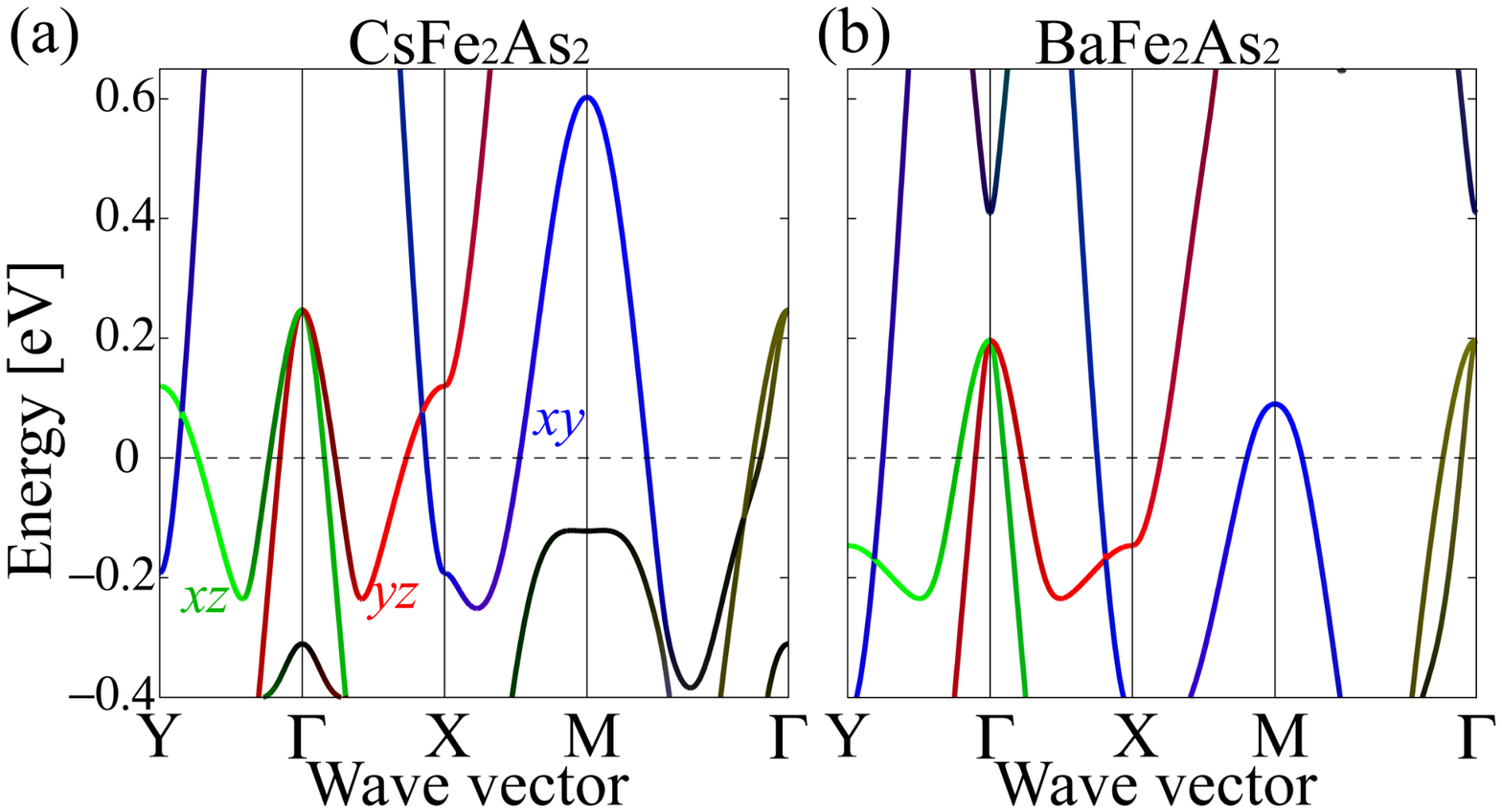}
\caption{
Bandstructures of the eight-orbital models for
(a) CsFe$_2$As$_2$ and (b) BaFe$_2$As$_2$.}
\label{fig:band}
\end{figure}

Here, we introduce the eight-orbital $d$-$p$ models for CsFe$_2$As$_2$
and BaFe$_2$As$_2$ analyzed in the main text.
We first derived the first principles tight-binding models
using the WIEN2k and WANNIER90 codes.
Next, we introduce the $k$-dependent energy shifts 
for orbital $l$, $\delta E_l$, 
by introducing the intra-orbital hopping parameters,
as we explain in Refs. \cite{S-FeSe-Yamakawa,S-FeSe-Onari}.
For the CsFe$_2$As$_2$ model,
we shift the $d_{xy}$-orbital band [$d_{xz/yz}$-orbital band] 
at ($\Gamma$, M, Y/X) points
by ($0$, $+0.4$, $0$) [($-0.4$, $0$, $+0.1$)] in unit eV.
For the BaFe$_2$As$_2$ model,
we do not introduce any energy shifts.
Figure \ref{fig:band} shows the bandstructures of 
the obtained (a) CsFe$_2$As$_2$ model and (b) BaFe$_2$As$_2$ model.

Next, we explain the multiorbital Coulomb interaction.
The bare Coulomb interaction for the spin channel 
in the main text is
\begin{equation}
(\Gamma^{\mathrm{s}})_{l_{1}l_{2},l_{3}l_{4}} = \begin{cases}
U_{l_1,l_1}, & l_1=l_2=l_3=l_4 \\
U_{l_1,l_2}' , & l_1=l_3 \neq l_2=l_4 \\
J_{l_1,l_3}, & l_1=l_2 \neq l_3=l_4 \\
J_{l_1,l_2}, & l_1=l_4 \neq l_2=l_3 \\
0 , & \mathrm{otherwise}.
\end{cases}
\end{equation}
Also, the bare Coulomb interaction for the charge channel is
\begin{equation}
({\hat \Gamma}^{\mathrm{c}})_{l_{1}l_{2},l_{3}l_{4}} = \begin{cases}
-U_{l_1,l_1}, & l_1=l_2=l_3=l_4 \\
U_{l_1,l_2}'-2J_{1_1,l_2} , & l_1=l_3 \neq l_2=l_4 \\
-2U_{l_1,l_3}' + J_{l_1,l_3} , & l_1=l_2 \neq l_3=l_4 \\
-J_{1_1,l_2} , &l_1=l_4 \neq l_2=l_3 \\
0 . & \mathrm{otherwise}.
\end{cases}
\end{equation}
Here, $U_{l,l}$, $U_{l,l'}'$ and $J_{l,l'}$
are the first principles Coulomb interaction terms for $d$-orbitals of BaFe$_2$As$_2$
given in Ref. \cite{S-Arita}.

Using the multiorbital Coulomb interaction,
the spin (charge) susceptibility in the RPA is given by  
\begin{equation}
{\hat \chi}^{s(c)}(q)={\hat\chi}(q)[1-{\hat \Gamma}^{s(c)}{\hat
\chi^0(q)}]^{-1},
\end{equation}
where irreducible susceptibilities is
\begin{equation}
\chi^0_{l,l';m,m'}(q)= -\frac{T}{N}\sum_k
G_{l,m}(k+q)G_{m',l'}(k).
\end{equation}
Here, ${\hat G}(k)$ is the multiorbital Green function 
introduced in the main text.
The $b$-channel interaction ($b=s,c$)
given by the RPA is 
$\hat{V}^{b}(q)=\hat{\Gamma}^{b}+\hat{\Gamma}^{b}\hat{\chi}^{b}(q)
\hat{\Gamma}^{b}$.

As shown by the Feynman
diagram in Fig. \ref{fig:FS} (e), $\hat{I}^{\bm{q}}(k,k')$ is given as
\begin{eqnarray}
&& \!\!\!\!\!\!\!\!\!\!\!
I^{\bm{q}}_{l,l';m,m'}(k,k')=\sum_{b=s,c}
\left[\frac{a^b}{2} V^{b}_{l,m;l',m'}(k-k')\right.
\nonumber \\
&&
-\frac{T}{N}\!\!\!\!\sum_{p,l_1,l_2,m_1,m_2}\!\!\!\!\!\!\!\!\!\!
 \frac{a^b}{2} V^{b}_{l,l_1;m,m_2}\left(p+\frac{\q}{2}\right)V^{b}_{m',l_2;l',m_1}\left(p-\frac{\q}{2}\right)
 \nonumber \\
&& \qquad\qquad
\times G_{l_1,m_1}(k-p)G_{l_2,m_2}(k'-p)
 \nonumber \\
&& 
-\frac{T}{N}\!\!\!\!\sum_{p,l_1,l_2,m_1,m_2}\!\!\!\!\!\!\!\!\!\!
 \frac{a^b}{2} V^{b}_{l,l_1;l_2,m'}\left(p+\frac{\q}{2}\right)V^{b}_{m_2,m;l',m_1}\left(p-\frac{\q}{2}\right)
 \nonumber \\
&& \qquad\qquad
\left.\times G_{l_1,m_1}(k-p)G_{l_2,m_2}(k'+p)\right],
\nonumber \\
\label{eqn:S-K} 
\end{eqnarray}
%
where $a^{s(c)}=3\ (1)$ and $p=(\p,\w_l)$.
In Eq. (\ref{eqn:S-K}),
the first line corresponds to the Maki-Thompson (MT) term,
and the second and the third lines give AL1 and AL2 terms, respectively.
Double-counting second-order terms with respect to
$\hat{\Gamma}^{s(c)}$ have to be subtracted.

\subsection{B: $\BI$ bond ordered state}
\begin{figure}[!htb]
\includegraphics[width=.99\linewidth]{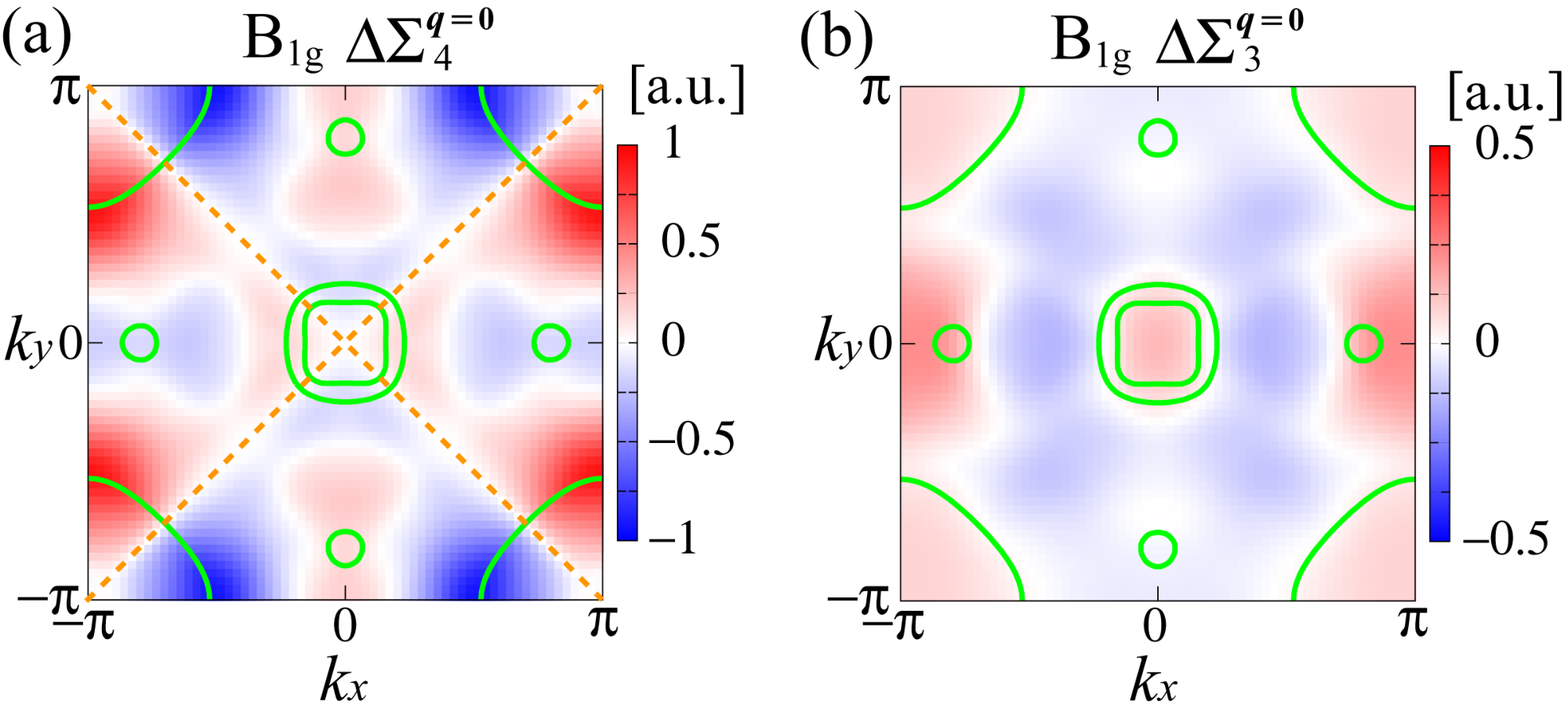}
\caption{
(a,b) $\BI$ symmetry form factors at $\q={\bm 0}$
obtained as the second largest eigenvalue.
The primary form factor on $d_{xy}$ orbital,
$\Delta\Sigma_{4}^{\bm{0}}\propto \cos k_x -\cos k_y$,
gives the nearest-neighbor bond order.
Orange dotted lines represent the symmetry nodes.
}
\label{fig:Sigma2}
\end{figure}

Figures \ref{fig:Sigma2}(a) and \ref{fig:Sigma2}(b) show 
the form factors for the second largest eigenvalue $\lambda=0.88$.
The obtained solution has $\BI$-symmetry
since the symmetry relation
$\Delta\Sigma^{\bm{0}}_{4}(k_x,k_y)\propto \cos k_x-\cos k_y$ holds.
This corresponds to the
nearest-neighbor bond order for $d_{xy}$ orbital.
This $\BI$ bond order induces
small secondary orbital order with $\BI$ symmetry as shown in
Fig. \ref{fig:FS} (a).

\subsection{C: Nematic susceptibility}
\begin{figure}[!htb]
\includegraphics[width=.99\linewidth]{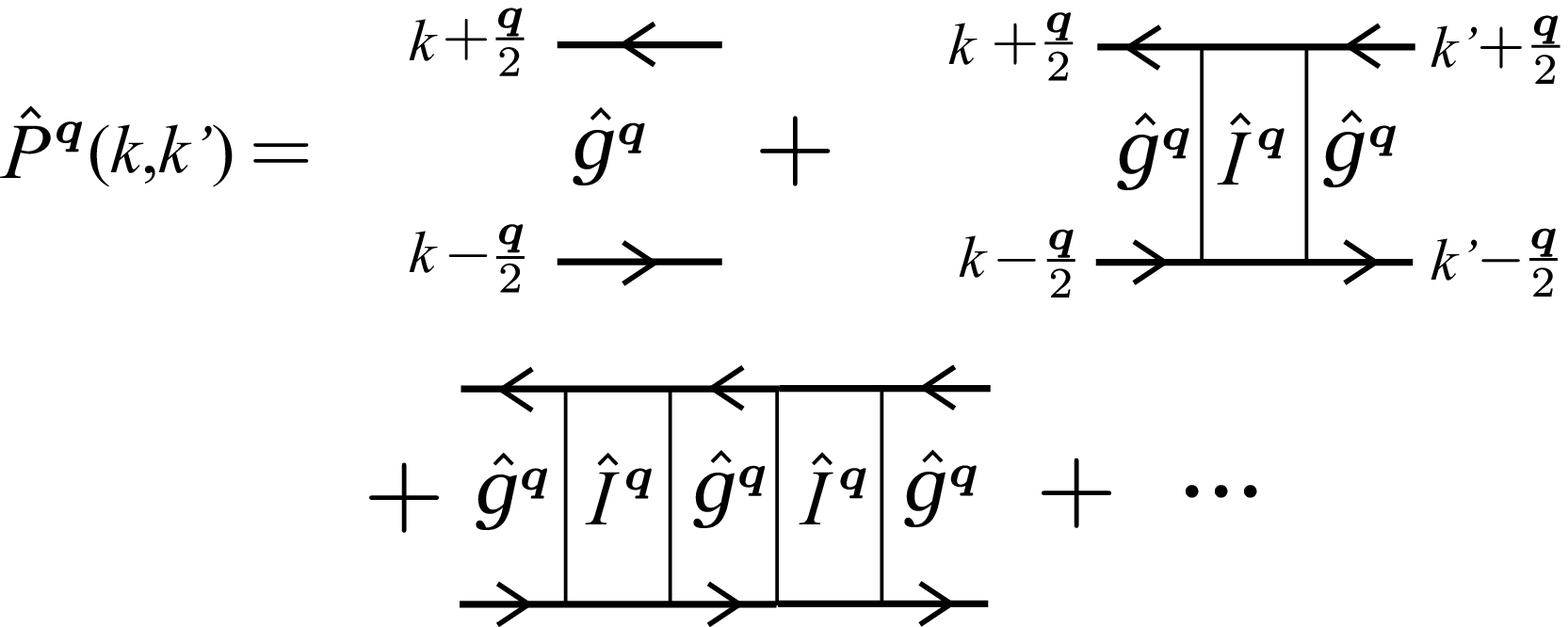}
\caption{
 Feynman diagrams of the two-particle Green function
$\hat{P}^{\bm{q}}(k,k')$.
The irreducible four-point vertex 
$\hat{I}^{\bm{q}}(k,k')$ is introduced in Fig. 1 (g) 
in the main text.
}
\label{fig:P}
\end{figure}

Next, we discuss the DW susceptibility with respect to
the form factor $\Delta\hat{\Sigma}$; ${\hat\chi}^{\Delta\Sigma}$.
By including both AL and MT vertex terms, it is given as
\begin{eqnarray}
\chi^{\Delta\Sigma}_{l,l';m,m'}(q)&=&-\frac{T^2}{N^2}\sum_{k,k'}\Delta\Sigma^{-\bm{q}}_{l,l'}(-\k)P^{\bm{q}}_{l,l';m,m'}(k,k')
\nonumber \\
& &\times \Delta\Sigma^{\bm{q}}_{m,m'}(\k'),
\end{eqnarray}
where
$\hat{P}^{\bm{q}}(k,k')=\hat{g}^{\bm{q}}(k)[\delta_{k,k'}+\hat{K}^{\bm{q}}(k,k')+\frac{T}{N}\sum_{k''}\hat{K}^{\bm{q}}(k,k'')\hat{K}^{\bm{q}}(k'',k')+\cdots]$. 
In Fig. \ref{fig:P},
we shown the Feynman diagram for $\hat{P}^{\bm{q}}(k,k')$,
in which higher-order MT and AL terms are included.
Using the Eq. (2), we can show that
\begin{eqnarray}
\hat{\chi}^{\Delta\Sigma}(q)=(1-\lambda_{\bm{q}})^{-1}\frac{-T}{N}
\sum_{k}\Delta\hat{\Sigma}^{-\bm{q}}(-\k)\hat{g}^{\bm{q}}(k)\Delta\hat{\Sigma}^{\bm{q}}(\k). 
\nonumber\\
\end{eqnarray}
Thus, the DW with wavevector $\q$ emerges 
when ${\hat\chi}^{\Delta\Sigma}(\q) \propto (1-\lambda_{\bm{q}})^{-1}$ diverges.

\subsection{D: Detailed explanation for the kernel function}
\begin{figure}[!htb]
\includegraphics[width=.99\linewidth]{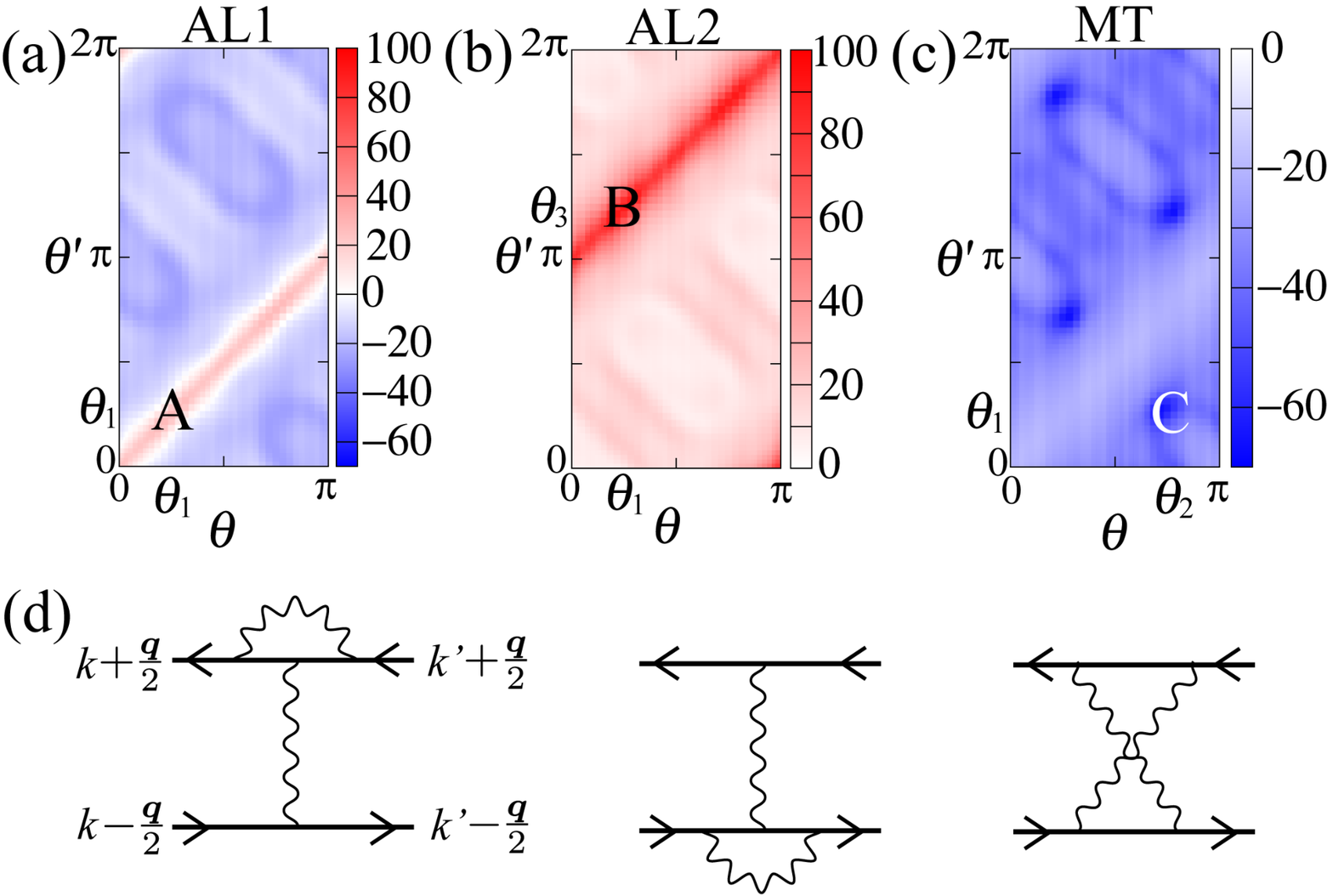}
\caption{
(a,b,c) $K_{\rm FS3}(\theta,\theta')$ given by the AL1 term, the AL2
 term, and the MT term, respectively.
(d) Second-order diagrams with respect to ${\hat \chi}(q)$
except for the AL terms. 
}
\label{fig:Kernel2}
\end{figure}

In Fig. \ref{fig:Kernel} (a) in the main text, we show the
momentum-dependence of the kernel function $K_{\rm FS3}(\theta,\theta')$
given by the all vertex terms.
Here, we discuss the contribution from each vertex term.
Figure \ref{fig:Kernel2} (a) shows $K_{\rm FS3}(\theta,\theta')$ given by
the AL1 term. The positive $K_{\rm FS3}(\theta,\theta')$ in the line region
$\theta'\sim\theta$ $({i.e.}, \ \k=\k')$ including the pair A 
comes from the particle-hole channel $\phi_{\mbox{p-h}}\equiv
T\sum_p^{|\w_l|<\w_{c}} G_{4,4}(k-p)G_{4,4}(k'-p)$.
Here, the cutoff energy $\w_c \ll E_{\rm F}$ corresponds to energy-scale of ${\hat \chi}^s$ in $\hat{V}^s$.
It is easy to show that 
$\phi_{\mbox{p-h}}$ takes large positive value for $\k=\k'$ in the case
of $\w_c \ll E_{\rm F}$.

Figure \ref{fig:Kernel2} (b) shows $K_{\rm FS3}(\theta,\theta')$ given by
the AL2 term. The positive $K_{\rm FS3}(\theta,\theta')$ in the line
region $\theta'=\theta+\pi$ $({i.e.}, \ \k=-\k')$ including the pair B
stems from the particle-particle (Cooper)
channel $\phi_{\rm p\mbox{-}p}\equiv T\sum_{p}^{|\w_l|<\w_{c}}
G_{4,4}(k-p)G_{4,4}(k'+p) \propto
\sum_{\bm{p}}\frac{1-f(\xi_{4}(\k-\bm{p}))-f(\xi_{4}(\k'+\bm{p}))}
{\xi_{4}(\k-\bm{p})+\xi_{4}(\k'+\bm{p})}$,
which diverges logarithmically for $\k=-\k'$ at $T=0$.
Here, $f(\e)$ is Fermi distribution function, 
$\xi_{4}(\k)$ is $d_{xy}$-orbital hole-band dispersion.

Figure \ref{fig:Kernel2}(c) shows 
$K_{\rm FS3}(\theta,\theta')$
given by the MT term.
The MT term assists the $\BII$ symmetry solution
since the negative value of $K_{\rm FS3}(\theta,\theta')$ is maximized
for the pair C$=(3\pi/4,\pi/4)$, as discussed in the main text.
However, the contribution of the MT term is smaller
than that of the AL terms as follows.
In fact, if we drop the AL terms in the DW equation,
the eigenvalue is quite small.

Here, we explain why the AL terms dominate over the MT term 
near the magnetic criticality based on the
spin fluctuation theories
\cite{S-TPSC,S-Scalapino,S-Monthoux,S-Moriya,S-Kontani-rev}.
The dynamical spin susceptibility is approximately expressed as
\begin{eqnarray}
\chi^s(\q,\w_l)= \frac{a\xi^2}{1+\xi^2(\q-\Q)^2+|\w_l|/\w_{\rm sf}}
\end{eqnarray}
where $\xi$ is the magnetic correlation length.
The relation $\xi^2\propto (T-T_N)^{-1}$ in the paramagnetic state
according to spin fluctuation theories.
$\w_{\rm sf} \propto \xi^{-2}$ is the energy scale of spin fluctuations.
Now, we discuss the absolute value of kernel in DW equation (2) 
in the main text, $f\equiv |T\sum_{k'} {\hat K}^{\bm{0}}(k,k')|$,
in the case of $\xi\gg1$ and $\w_{\rm sf}\ll 2\pi T$.
When the kernel for $\q={\bm0}$ is given by the AL term,
$f_{\rm AL}\sim \sum_\p\{\chi^s(\p,0)\}^2\sim \xi^2$ 
in two-dimensional systems at a fixed $T$.
(The electron Green functions in the AL diagram also give
important $T$-dependence as discussed in 
Refs. \cite{S-FeSe-Yamakawa,S-FeSe-Onari}.)
When the kernel is given by the MT term,
$f_{\rm MT}\sim \sum_\p\chi^s(\p,0)\sim \log\xi$.
Therefore, the AL term dominates over the MT term
when $\xi\gg1$.
In the same way, the second-order diagrams except for the AL terms,
shown in Fig. \ref{fig:Kernel2} (d),
are scaled as $(\log\xi)^2$.
Therefore, the AL terms are the most important for $\xi\gg1$.
The significance of the AL terms
near the magnetic criticality
is verified by the functional-renormalization-group (fRG)
study in Refs. \cite{S-Tsuchiizu-Cu,S-Tsuchiizu-CDW,S-Tsuchiizu-Ru1}.

When $U$ is small and the relation  
$V^s(q)=U+U\chi^s(q)U \sim U$ holds, 
the AL term is very small and impossible to stabilize the bond order. 
With increasing $U$, the AL term becomes large in the case that 
$\chi^s(\p,\w_l)$ strongly develops for $\p\sim \Q$ 
at low energies.
In the present model for AFe$_2$As$_2$, 
moderate spin fluctuations (Stoner factor $\a_S\approx 0.93$) 
are requied for the AL-term driving nematic order, 
whereas much weaker spin fluctuations are enough for FeSe 
as discussed in Refs. \cite{S-FeSe-Yamakawa,S-FeSe-Onari}.

\subsection{E: Origin of $B_{1g}$ symmetry orbital order in undoped compounds
}

\begin{figure}[!htb]
\includegraphics[width=.7\linewidth]{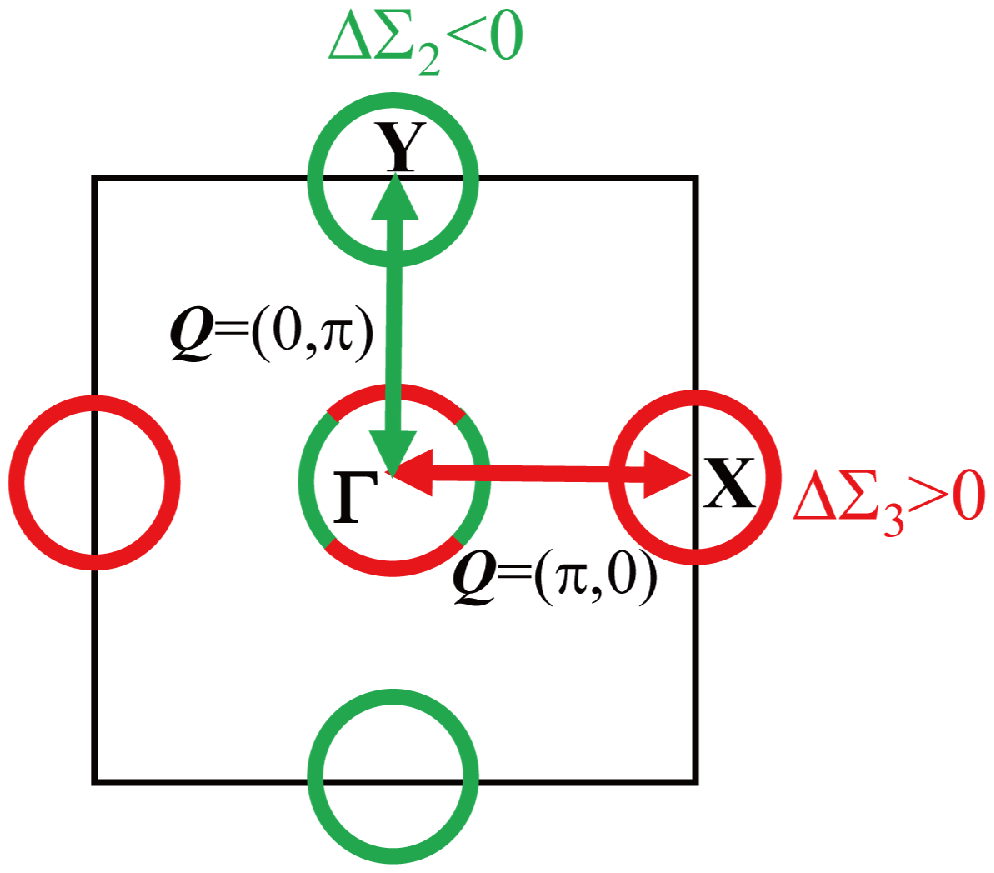}
\caption{
Simplified FSs for $n_d\approx 6$ composed of only 
$d_{xz}$ and $d_{yz}$ orbitals.
Due to the intra-orbital FS nesting,
spin fluctuations on $d_{xz}$ [$d_{yz}$] orbital develop 
at $\Q=(0,\pi)\ [\Q=(\pi,0)]$.
These spin fluctuations induce nematic orbital order
($n_{xz}\ne n_{yz}$) cooperatively.
}
\label{fig:S-simpleFS}
\end{figure}

Here, we briefly explain the reason why 
$\BI$ symmetry orbital order ($n_{xz}\ne n_{yz}$) appears
in usual undoped ($n_d=6$) Fe-based superconductors.
Figure \ref{fig:S-simpleFS} shows a simplified FSs,
in which only $d_{xz}$ and $d_{yz}$ orbitals are shown.
Here, spin fluctuations on $d_{xz[yz]}$-orbital develop 
at wavevector $\Q=(0,\pi)\ [(\pi,0)]$,
due to the good intra-orbital FS nesting.

Here, we consider the DW equation (2) at $\q=\bm{0}$.
In the kernel for $d_{yz}$ orbital, $K_{3,3;3,3}^{\bm{0}}(\k,\k')$,
the AL terms give large positive value for 
$\k,\k'\approx \k_{X}$ or $\k_{\Gamma}$.
In contrast, the MT term give negative contribution
for $\k\approx \k_{X}$ and $\k'\approx \k_{\Gamma}$.
Therefore, the form factor
$\Delta\Sigma^{\bm{0}}_{3,3}(\k)$
takes large value in magnitude for $\k\approx \k_{X},\k_{\Gamma}$.
($\Sigma^{\bm{0}}_{3,3}(\k)|$ may have sign reversal
between $\Gamma$ and $X$ points due to the MT term.)
In the same way,
$|\Delta\Sigma^{\bm{0}}_{2,2}(\k)|$
takes large value for $\k\approx \k_{Y},\k_{\Gamma}$.

In the Hubbard model,
the net charge density (=charge monopole) order is 
strongly suppressed by the on-site Coulomb interaction $U$.
In contrast, both the orbital order and the bond order can appear
since they are (non-local) charge quadrupole orders.
For this reason, the relation 
$\Delta\Sigma^{\bm{0}}_{2,2}(k_x,k_y)=
-\Delta\Sigma^{\bm{0}}_{3,3}(k_y,k_x)$
is satisfied by solving the DW equation (2).
This solution gives the orbital order ($n_{xz}\ne n_{yz}$)
without net charge density modulation.
Thus, spin fluctuations on $d_{xz}$ and $d_{yz}$ orbitals
induce the orbital order ($n_{xz}\ne n_{yz}$) cooperatively.
More detailed explanation is given in 
Refs. \cite{S-FeSe-Yamakawa,S-FeSe-Onari}.

Thus, the present study reveals the significant
roles of FS orbital character and FS topology 
on the nature of nematicity.
In usual compounds ($n_d\sim6$) with FSs in Fig. \ref{fig:S-simpleFS},
spin fluctuations on ($d_{xz},d_{yz}$) orbitals strongly develop.
In this case, the nematic orbital order naturally appears.
In heavily hole-doped compounds ($n_d\sim5.5$),
spin fluctuations develop solely in $d_{xy}$ orbital.
Even in this case, nematic transition can appear by forming 
the bond order spontaneously
as we revealed in the main text.
We comment that the symmetry of nematicity is not simply related to 
the direction of wavevector of spin fluctuations.
In summary, the diverse nematicity in Fe-based superconductors
(such as $B_{1g}$ orbital order and $B_{2g}$ bond order)
originates from the rich compound dependence of 
FS orbital character and FS topology.


\subsection{F: Conserving approximation}
In the main text, the self-energy correction is not included 
in the kernel function $K$.
For this reason, the DW equation in the main text
does not satisfy the condition of the conserving approximation 
(CA) formulated by Baym and Kadanoff.
The great merit of the CA is that the 
macroscopic conservation laws are satisfied rigorously.
This merit is important to avoid unphysical results.
Here, we first calculate the one-loop self-energy using 
the fluctuation exchange (FLEX) approximation \cite{S-FLEX,S-Onari-form}.
Next, we analyze the DW equation with including the FLEX
self-energy, in order to satisfy the criteria of the CA.

\begin{figure}[!htb]
\includegraphics[width=.99\linewidth]{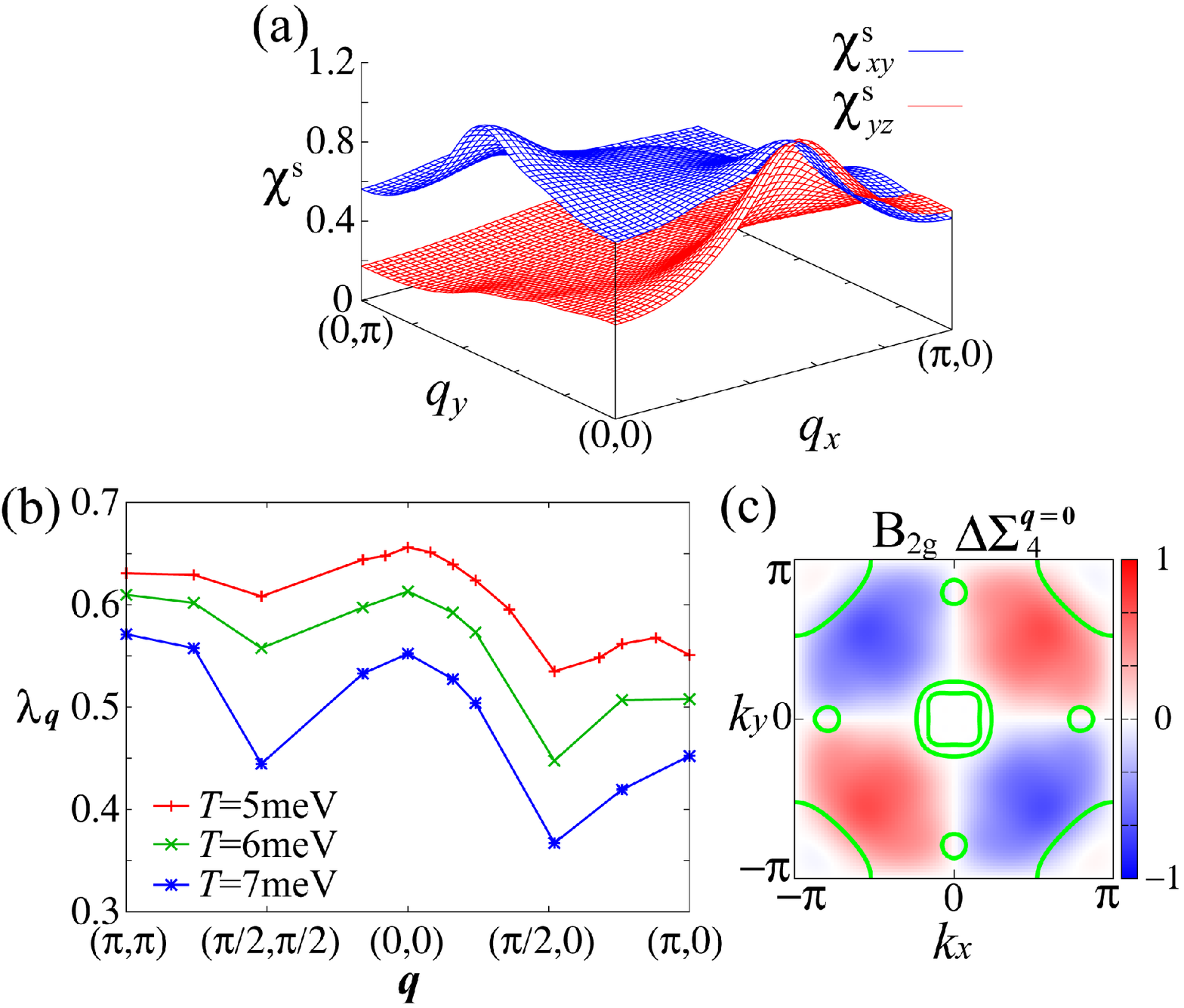}
\caption{
(a) $\q$ dependences of 
 $\chi^{s}_{xy}(\q,0)$ and $\chi^{s}_{yz}(\q,0)$ given by the FLEX approximation.
(b) $\bm{q}$ dependences of the maximum eigenvalue 
obtained by the present improved linearized DW equation.
(c) Obtained form factors at $\q={\bm 0}$ with $\BII$ symmetry.
}
\label{fig:FLEX}
\end{figure}

The FLEX self-energy (with $C_4$ symmetry) is given by 
$\hat{\Sigma}(k)=\frac{T}{N}\sum_q\hat{V}^\Sigma(q)\hat{G}(k-q)$,
where
${\hat G}(k)=[(i\e_n-\mu){\hat1}-{\hat{h}}^0(\k)-\hat{\Sigma}(k)]^{-1}$ 
is the Green function with the self-energy,
and $\hat{V}^\Sigma$ given as
$\displaystyle 
\frac32 {\hat \Gamma}^s{\hat \chi^s}(q){\hat \Gamma}^s
+\frac12 {\hat \Gamma}^c{\hat \chi^c}(q){\hat \Gamma}^c
-\frac12 \bigl[{\hat \Gamma}^c{\hat\chi}^0(q){\hat \Gamma}^c
+{\hat \Gamma}^s{\hat\chi}^0(q){\hat \Gamma}^s
-\frac14 ({\hat \Gamma}^s+{\hat \Gamma}^c){\hat\chi}^0(q)
({\hat \Gamma}^s+{\hat \Gamma}^c) \bigr]$.
We solve $\hat{\Sigma}$, $\hat{G}$, and $\hat{\chi}^{s(c)}$
self-consistently. 
Figure \ref{fig:FLEX} (a) shows the $\q$-dependence of $\chi^s_{xy(yz)}$
given by the FLEX approximation for
$T=5\sim7$meV at fixed $r=0.96$ ($\a_S=0.93$ at $T=5$meV) 
by employing $N=100\times100$ $\k$-meshes. 
The obtained $\q$-dependence of $\chi^s_{xy(yz)}$ 
in the FLEX approximation is similar to that given by the RPA
in the main text.

Next, we construct the improved DW equation 
to satisfy the framework of the CA,
by introducing the obtained $\hat{\Sigma}$ and ${\hat \chi}^{s,c}$ 
into Eqs. (2)-(3) in the main text.
Figure \ref{fig:FLEX} (b) shows the eigenvalue $\lambda_{\q}$
given by solving the improved DW equation.
It is confirmed that $\lambda_{\q}$ shows the maximum 
at $\q=\bm{0}$ for $T=5$meV($\lambda_{\bm{0}}=0.66$), 
consistently with the result without the 
self-energy in Fig. 2 (d).
The obtained form factor $\Delta\Sigma_{4}^{\bm{0}}$ 
has $\BII$ symmetry as shown in Fig. \ref{fig:FLEX} (c), which 
is similar to Fig. 2 (a) in the main text.
Thus, the results in the main text are verified 
by the present improved DW equation 
that satisfies the condition of the CA.
The magnitude of $\lambda_{\q}$ is 
suppressed by including the self-energy.
Although we cannot calculate for $T<5$meV 
due to the lack of frequency- and $\k$-mesh numbers,
the value of $\lambda_{\q}$ will reach unity at lower temperature.



\begin{thebibliography}{99}
\bibitem{conserving}
The one-loop (single-fluctuation-exchange) self-energy 
is given by
$\hat{\Sigma}(k)=\frac{T}{N}\sum_q\hat{V}^\Sigma(q)\hat{G}(k-q)$, 
where
$\displaystyle 
\hat{V}^\Sigma = \frac32 {\hat \Gamma}^s{\hat \chi^s}(q){\hat \Gamma}^s
+\frac12 {\hat \Gamma}^c{\hat \chi^c}(q){\hat \Gamma}^c
-\frac12 \bigl[{\hat \Gamma}^c{\hat\chi}^0(q){\hat \Gamma}^c
+{\hat \Gamma}^s{\hat\chi}^0(q){\hat \Gamma}^s
-\frac14 ({\hat \Gamma}^s+{\hat \Gamma}^c){\hat\chi}^0(q)
({\hat \Gamma}^s+{\hat \Gamma}^c) \bigr]$.
\end{thebibliography}
\end{document}